\documentclass[preprint]{ptephy_v1}
\preprintnumber{XXXX-XXXX} 

\usepackage{url} 

\begin{document}
\title{Overview of  KAGRA : Detector design and construction history
}

\author[1,2]{T.~Akutsu}
\author[3,4,1]{M.~Ando}
\author[5]{K.~Arai}
\author[5]{Y.~Arai}
\author[6]{S.~Araki}
\author[7]{A.~Araya}
\author[3]{N.~Aritomi}
\author[8,9]{Y.~Aso}
\author[10]{S.~Bae}
\author[11]{Y.~Bae}
% ...
\affil[1]{Gravitational Wave Project Office, National Astronomical Observatory of Japan (NAOJ), Mitaka City, Tokyo 181-8588, Japan}
\affil[2]{Advanced Technology Center, National Astronomical Observatory of Japan (NAOJ), Japan}
\affil[3]{Department of Physics, The University of Tokyo, Bunkyo-ku, Tokyo 113-0033, Japan}
\affil[4]{Research Center for the Early Universe (RESCEU), The University of Tokyo, Bunkyo-ku, Tokyo 113-0033, Japan}
\affil[5]{Institute for Cosmic Ray Research (ICRR), KAGRA Observatory, The University of Tokyo, Kashiwa City, Chiba 277-8582, Japan}
\affil[6]{Accelerator Laboratory, High Energy Accelerator Research Organization (KEK), Tsukuba City, Ibaraki 305-0801, Japan}
\affil[7]{Earthquake Research Institute, The University of Tokyo, Bunkyo-ku, Tokyo 113-0032, Japan}
\affil[8]{Kamioka Branch, National Astronomical Observatory of Japan (NAOJ), Kamioka-cho, Hida City, Gifu 506-1205, Japan}
\affil[9]{The Graduate University for Advanced Studies (SOKENDAI), Mitaka City, Tokyo 181-8588, Japan}
\affil[10]{Korea Institute of Science and Technology Information (KISTI), Yuseong-gu, Daejeon 34141, Korea}
\affil[11]{National Institute for Mathematical Sciences, Daejeon 34047, Korea}
% ...
\affil[\space]{(KAGRA Collaboration)}

\date{\today}
\begin{abstract}%
KAGRA is a newly built gravitational-wave telescope, a laser interferometer comprising arms with a length of 3\,km, located in Kamioka, Gifu, Japan. 
KAGRA was constructed under the ground and it is operated using cryogenic mirrors that help in reducing the seismic and thermal noise.  Both technologies are expected to provide directions for the future of gravitational-wave telescopes.  In 2019, KAGRA finished all installations with the designed configuration, which we call {\it the baseline KAGRA}. 
In this occasion, we present an overview of the baseline KAGRA from various viewpoints in a series of of articles. In this article, we introduce the design configurations of KAGRA with its historical background. 
\end{abstract}


\maketitle
\section{Introduction}\label{ptep01_sec1}

KAGRA is the world's first km-scale cryogenic gravitational-wave telescope built under the ground in the Kamioka mine; it has been successfully constructed and has 
 performed its observation runs, the latest of which overlapped with that of
so-called second-generation telescopes such as Advanced LIGO~\cite{aligo} in the US and Advanced Virgo~\cite{adV} in Italy.
KAGRA implements two key technologies that are important for so-called third-generation telescopes in the
future like Einstein Telescope~\cite{ET} and Cosmic Explorer~\cite{CE}. One is to build entire facility under the ground to reduce the seismic noise and the seismic Newtonian noise. The other is to cool the mirrors down to the cryogenic temperature to reduce the mirror thermal noise, which is the limiting noise source (at approximately 100\,Hz) for other telescopes operating at the room temperature ($\sim$300\,K). 
Because of this KAGRA can be categorized as a 2.5 generation gravitational-wave telescope.

KAGRA is designed to observe the gravitational waves from neutron star binaries at a distance of $\sim150\,\mathrm{Mpc}$ from the Earth. Although it would require a few more years to reach the design sensitivity, 
we decided to start the observing run in 2020 before the termination of the joint observing run by LIGO and Virgo (O3) with a lower sensitivity.
Along with the room-temperature operation in 2016 (the {\it iKAGRA} operation)~\cite{iKAGRA:PTEP} and the cryogenic test operation in 2018 (the {\it bKAGRA Phase-1} operation)~\cite{bKAGRAphase1}, 
this first observing run is an important milestone for the telescope.

In 2019, the installation works for the designed configuration of KAGRA were mostly finished. After a year-long interferometer commissioning, in February 2020, KAGRA finally started its first observing run and terminated it in April 2020. During this period, we had a month-long commissioning break to further improve the interferometer performance. Advanced LIGO and Advanced Virgo jointly started the O3 observing run in April 2019 and terminated the run in March 2020. The telescopes are planning another observing run, namely O4, after approximately 1.5 years of commissioning and upgrade works. The observing scenario for the telescopes is summarized in Ref.~\cite{ObservingScenario}.

This paper describes the design of the KAGRA telescope and summarizes the operations including our first observing run. The structure of the paper is as follows. In Section~\ref{ptep01_sec1_5}, we present an overview of the telescope. Section~\ref{ptep01_sec2} explains the fundamental noise sources of the telescope, namely quantum noise, thermal noise, and seismic ($+$ seismic Newtonian) noise. 
In Section~\ref{ptep01_sec4}, we review the past test operations and report our first observing run. Section~\ref{ptep01_sec5} summarizes the paper.

\section{Detector configuration}\label{ptep01_sec1_5}

\begin{figure}[htbp]
	\begin{center}
		\includegraphics[width=15cm]{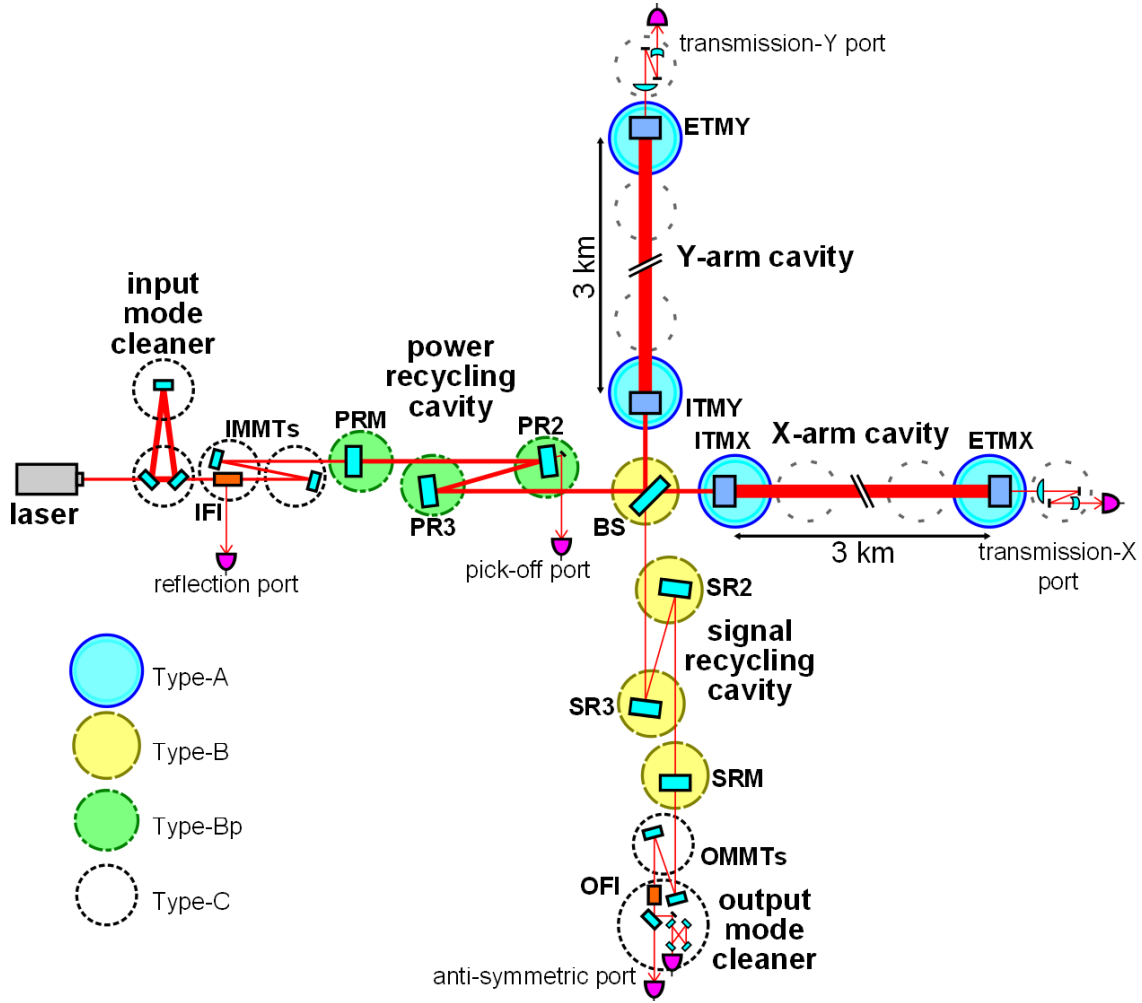}
	\caption{Schematic of the KAGRA interferometer. All mirrors with labels are suspended inside the vacuum tanks with four types of vibration isolation systems. Different types of circles in the figure represent different types of vibration isolation system. Vacuum tanks in front of the input and end test masses (depicted as dotted grey circles) contain narrow angle baffles and optical systems for photon calibrator. ITMX (Y): input test mass X (Y), ETMX (Y): end test mass X (Y), BS: beam splitter, PRM: power recycling mirror, SRM: signal recycling mirror, IMMT (OMMT): input (output) mode-matching telescope, IFI (OFI): input (output) Faraday isolator.}
	\label{fig:IFO}
	\end{center}
\end{figure}

\subsection{Overview of interferometer}
Figure~\ref{fig:IFO} shows a schematic of the KAGRA telescope that is based on a Michelson interferometer and has a Fabry-Perot cavity of 3\,km in each arm. The 3\,km arm cavities are formed by cryogenic sapphire mirrors and the other room temperature mirrors are made of fused silica. The interferometer is operated in the dark fringe; most of the input beam is reflected back to the direction of the laser source and only a small fraction of the light leaks to the signal extraction port owing to some asymmetry. A power recycling mirror is located between the interferometer and laser source such that the reflected light is re-injected to enhance the effective power of the interferometer~\cite{Meers}. Additionaly, a signal recycling mirror is located at the anti-symmetric port such that the frequency response of the interferometer can be selected by changing the resonant condition of the signal field in the signal recycling cavity~\cite{Meers}. The interferometer configuration is called Resonant Sideband Extraction, or RSE, in the case that the signal recycling cavity is set anti-resonance of the carrier light so that the detector bandwidth is broaden~\cite{Mizuno}. Each recycling cavity consists of a recycling mirror and two folding mirrors to increase the Gouy phase inside the recycling cavity. The power and alignment of the transmitted beam of each arm cavity are monitored using a transmission monitor system, which consists of beam reducing telescopes on a vibration isolated table to reduce the size of the beam. The input and output mode cleaners are utilized to cut off the unnecessary light. 
The gravitational wave signal is extracted from the transmitted power of the output mode cleaner and sent to the digital control system.

\subsection{Facility and vacuum system}

The walls of the underground facility are coated with anti-dust paint and the entire facility complies with the ISO class 6 cleanroom standards~\cite{ISO}. The laser room and the clean booths for the cryostats follow the ISO class 1 cleanroom standards and the clean booths for other vacuum chambers follow the ISO class 4 cleanroom standards. The ceilings are covered with plastic sheets to protect the facility from the underground water drops. Each arm tunnel with a height and width of 4\,m each accommodates a 3\,km KAGRA vacuum duct ($\phi800\,\mathrm{mm}$), and a ventilation duct to source clean fresh air from the outside. The KAGRA vacuum duct is located 0.5\,m off the center of the tunnel to alllow an electric motor car to pass through the tunnel. A 1.5\,km vacuum duct for the geophysics interferometer~\cite{GIF} ($\phi400\,\mathrm{mm}$) is along X-arm cavity. A $\phi400\,\mathrm{mm}$ water drainage pipe is located underneath the concrete floor of the tunnel and two additional drainage pipes are located in the Y-arm tunnel.

The ducts cannot be baked after the installation owing to the mining law but ultra-high vacuum of the order $10^{-7}$\,Pa can be realized using the electro-chemical buffing technique. Each 3\,km duct consists of 250 pieces of 12\,m tube that were connected inside the tunnel using a mobile clean booth. Each flange of the 12\,m tube accommodates a duct baffle made of bright-annealed stainless steel and coated with a black coating whose reflectivity for the 1064\,nm light is as low as 3\,\%~\cite{AkutsuOME}.

The vacuum chambers are set on a layer of 20\,mm thick flat mortar and anchored on the bedrock  covered with $\sim200$\,mm thick concrete. The flatness of the mortar is less than 2\,mm. Each vacuum chamber is set within 2\,mm from the marking point measured inside the mine with an accuracy of approximately $2\sim3$\,mm.


\begin{figure}[htbp]
	\begin{center}
		\includegraphics[width=15cm]{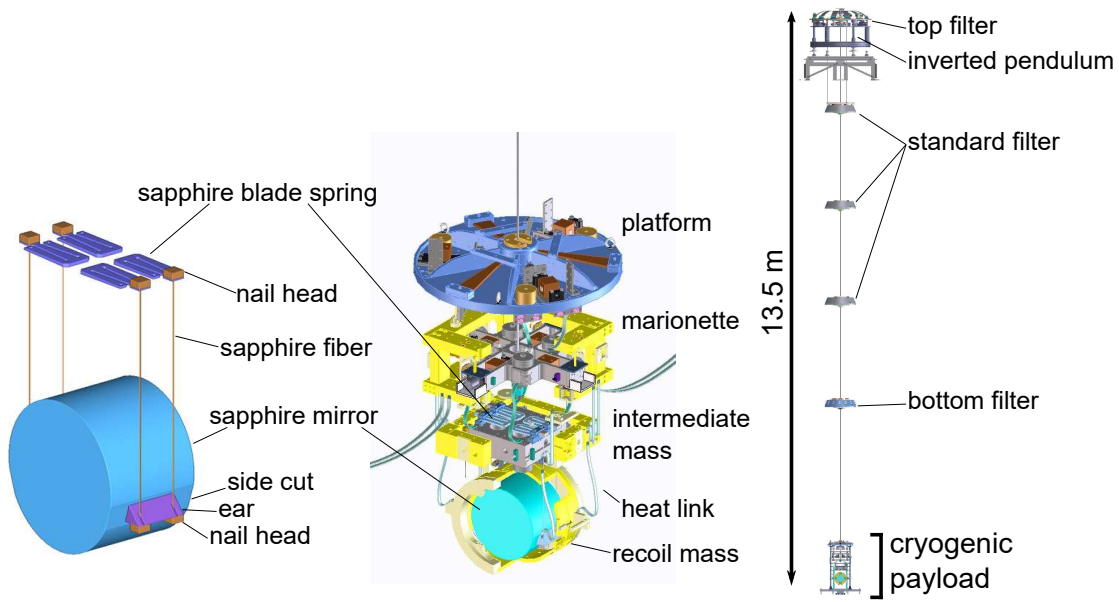}
	\caption{{\it Left}: Sapphire suspension. {\it Center}: Cryogenic payload. The recoil masses for each stage are drawn in yellow, cut open so that the test mass chain inside is easier to see.  {\it Right}: Type-A seismic isolation systems.}
	\label{fig:isolation}
	\end{center}
\end{figure}

\subsection{Sapphire test mass and cryogenic payload}
The substrate of the test masses in the arm cavities is a c-axis sapphire. It has a diameter and thickness of 22 and 15\,cm, respectively, and a mass of 22.8\,kg.
The front surface of the substrate has a curvature of 1.9\,km and the back surface is flat with a 0.025\,$^\circ$ wedge in the horizontal direction for the input mirror and a 0.05\,$^\circ$ wedge for the end mirror. The surface figure error of the front surface was measured to be approximately 0.3 and 0.5\,nm before and after the coating, respectively. The high-reflective surface of the mirror is coated with silica/tantala doublets. The optical absorption in the coating of the input and end mirrors was measured to be less than 0.3 and 0.5\,ppm, respectively~\cite{HiroseMirror}. The optical absorption in the substrate of the input and end mirrors was measured to be below $50\,\mathrm{ppm}/\mathrm{cm}$ and $100\,\mathrm{ppm}/\mathrm{cm}$, respectively~\cite{HiroseMirror}. 

These sapphire mirrors are suspended as shown in the left panel of Fig.~\ref{fig:isolation}~\cite{Ushiba}. The "ear" in the figure is also made of sapphire and is attached to the $36\,\mathrm{mm}\times150\,\mathrm{mm}$ flat area on the mirror barrel surface via the hydro-catalysis bonding technique~\cite{HaughianHCB1,HaughianHCB2}. Each mirror is suspended by four sapphire fibers; these fibers are 35\,cm long and 1.6\,mm thick with a cubic nail head at each end. The upper surface of the nail head at the bottom of the fiber is attached to the bottom of the ear with a gallium foil, which is used for bonding. The lower surface of the nail head at the top of the fiber is attached to a sapphire blade. 
Because sapphire fibers are considerably thick, the stretch caused by the weight of the sapphire mirror is shorter than the difference between the length of fibers. To compensate this difference, sapphire blade springs were introduced.
The sapphire blade springs are fixed to the intermediate mass, which is cooled down to $\sim$16\,K. The transferable heat from the sapphire mirrors at temperature $T=T_1$ to the intermediate mass at $T=T_2$ is given by
\begin{eqnarray}
K&=&\int_{T_2}^{T_1}\frac{\pi d_{\rm fiber}^2}{4\ell_{\rm fiber}}\kappa_{\rm fiber}(T) dT
\end{eqnarray} 
where $d_{\rm fiber}$, $\ell_{\rm fiber}$, and $\kappa_{\rm fiber}$ denote the thickness, length, and thermal conductivity of the sapphire fiber, respectively. The temperature dependence of the thermal conductivity (almost proportional to the cube of temperature) should be considered. The thermal resistance of the bonds and blades should be included in the calculation. Consequently, the sapphire suspension fiber can transfer $\sim1$\,W heat from the mirror~\cite{SaschaCQG}. The input laser power is limited by the heat absorption and cooling capability to maintain the test mass operating temperature below 22\,K~{\cite{SaschaCQG}\cite{MichimuraPSO}}.

\begin{figure}[htbp]
	\begin{center}
		\includegraphics[width=15cm]{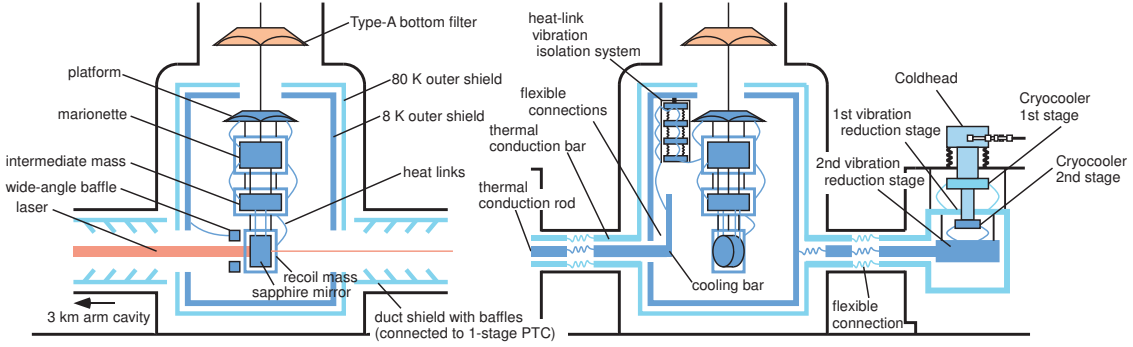}
	\caption{Schematic of the cooling system. Side view of the cryostat from the direction orthogonal to the arm ({\it left}) and another side view at an angle of 30\,$^\circ$ from direction indicated in the left panel ({\it right}). Although we omitted, there is another cryocooler on the left hand side of the right panel to cool down the cooling bar and cryogenic payload.}
	\label{fig:cryostat}
	\end{center}
\end{figure}

\subsection{Cryogenic system}
As depicted in the center panel of Fig.~\ref{fig:isolation}, the semi-monolithic sapphire suspension system described above constitutes the bottom part of the 4-stage cryogenic payload (platform, marionette, intermediate mass, and sapphire mirror) that consists of the passive vibration isolation system and the active control system to adjust the position and alignment of the sapphire mirror~\cite{Ushiba}. 
A recoil mass for each stage below the platform is located to mount the actuators. To cool these cryogenic payloads, 6\,N (99.9999\% purity) aluminum heat links are used. Each heat link is a strand of 49 wires (0.15\,mm in diameter) with a total outer diameter of $\sim1$\,mm. Although the measured resonant frequency of the stranded wire is as low as $\sim10$\,Hz, the vibration from the cryo-cooler could affect the sensitivity at low frequencies~\cite{Chen}; therefore, the heat links are attached to the marionette recoil mass of the cryogenic payload from the 3-stage vibration isolation system for heat links, as shown in the right panel of Fig.~\ref{fig:cryostat}.~\cite{Yamada}.

The cryogenic payloads are accommodated inside the 8\,K inner radiation shield and the 80\,K outer radiation shield in the cryostat vacuum chamber to be isolated from the 300\,K radiation from the environment(The details of the cryostat are described in Ref.~\cite{Sakakibara-review}). The radiation shield has an aperture that sends the beam to the sapphire mirror; moreover, the power of the 300\,K radiation through the aperture is as high as $\sim20$\,W. To terminate the invasion of the 300\,K radiation, a 5\,m long cryogenic duct-shield is located next to each end of the radiation shield along the beam axis. A black coating is applied to the inner surface of the cryogenic duct-shield to increase the absorption of the 300\,K radiation. Several baffles with black coating are equipped inside the cryogenic duct-shield to efficiently block the propagation of the 300\,K radiation. Consequently, the power of the 300\,K radiation invading through the radiation shield is less than 0.1\,W~\cite{SakakibaraCQG1,SakakibaraCQG2}. 
The baffle in the cryogenic duct-shield also helps in reducing the influence of the laser light scattered from the sapphire mirror at a small angle. In addition, we installed a cryogenic suspended baffle next to the sapphire mirror to block the laser light scattered from the sapphire mirror at a large angle. There are also room temperature baffles specially made to block the laser light scattered from the sapphire mirror in a small angle. They are placed in the chambers 36~m away from ETMs and 28~m away from ITMs.

Each cryostat accommodates four 2-stage low-vibration pulse tube cryo-coolers to cool down the cryogenic payload and radiation shields; additionally, single-stage pulse tube cryo-coolers are used to cool the duct-shields. 
In the first stage of the four 2-stage cryo-coolers (at approximately 65\,K), heat is extracted from the 80\,K outer shields. In the second stage of two of the 2-stage cryo-coolers (at approximately 8\,K), heat is extracted from the inner shields, and in the second stage of the remaining two 2-stage cryo-coolers, heat is extracted from the cryogenic payload using two cooling bars and the heat-link vibration isolation systems. 
Using this scheme, the cryogenic payload is thermally isolated from the radiation shield; the temperature of the test mass does not increase when large-angle scatterings from the mirror surface are absorbed by the radiation shield~\cite{Tokoku}. 
The temperature of the single-stage cryo-cooler for the cryogenic duct-shields is approximately 50\,K.

The pulse tube cryo-coolers for the payloads and shields are mounted on a vibration reduction system~\cite{Tokoku}.
The vibration reduction stage is rigidly connected to the ground to isolate the vibrations of the cold head and it is thermally connected to the cold head through soft copper heat links. The single-stage cryo-cooler for the duct-shield has a similar system.

\subsection{Seismic isolation system}
\begin{figure}[t]
	\begin{center}
		\includegraphics[width=5cm]{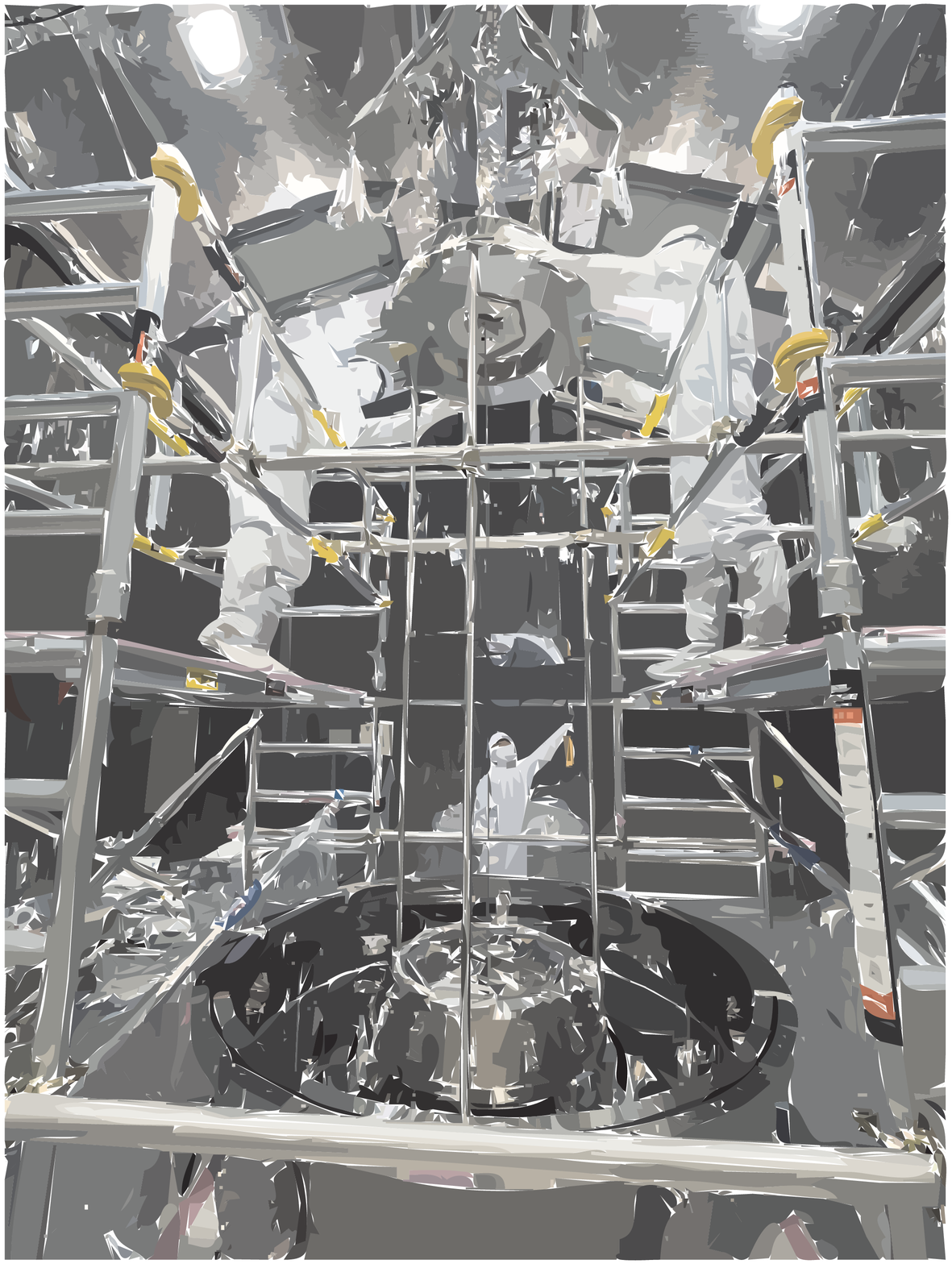}
		\includegraphics[width=5cm]{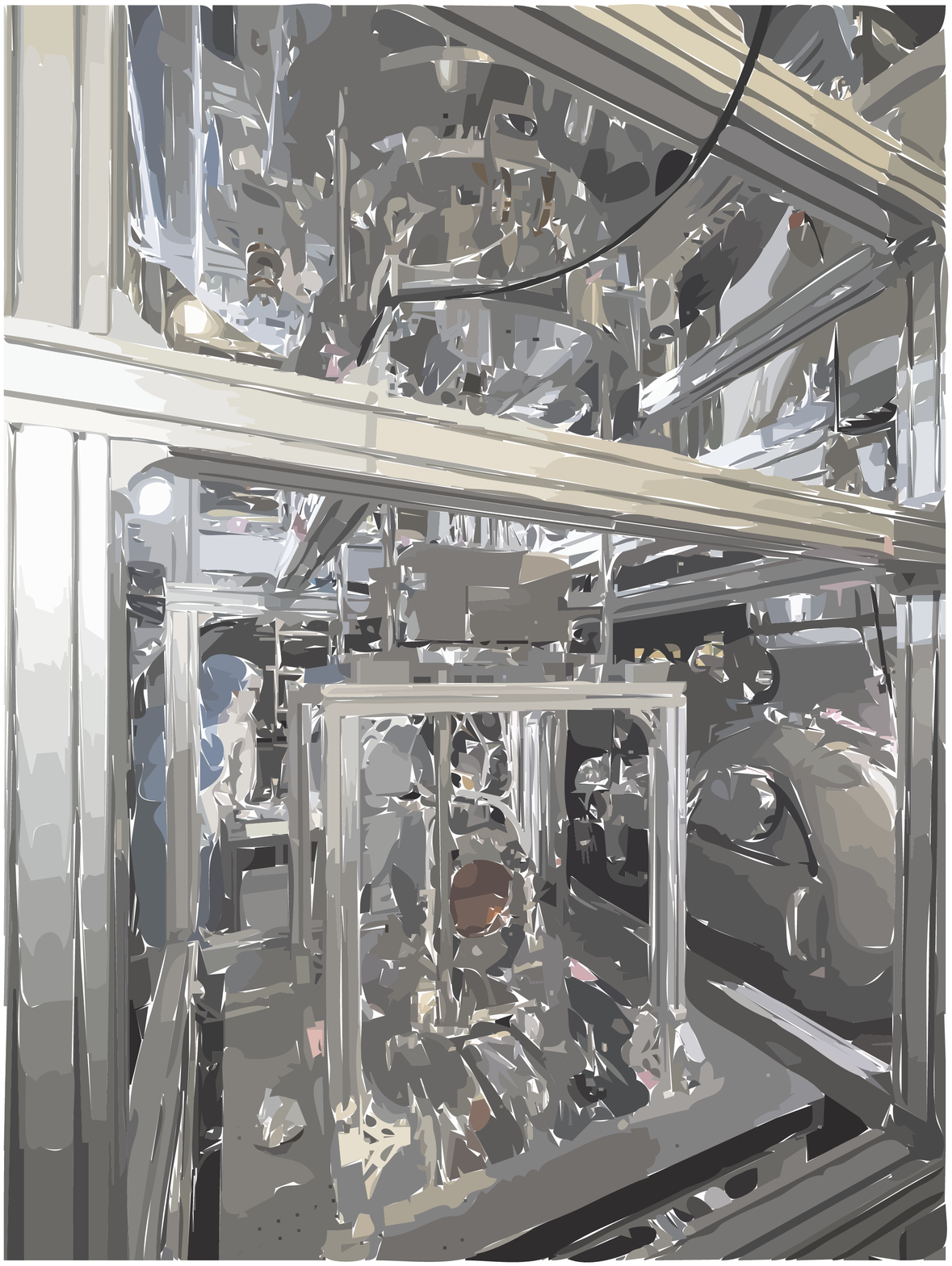}
		\includegraphics[width=5cm]{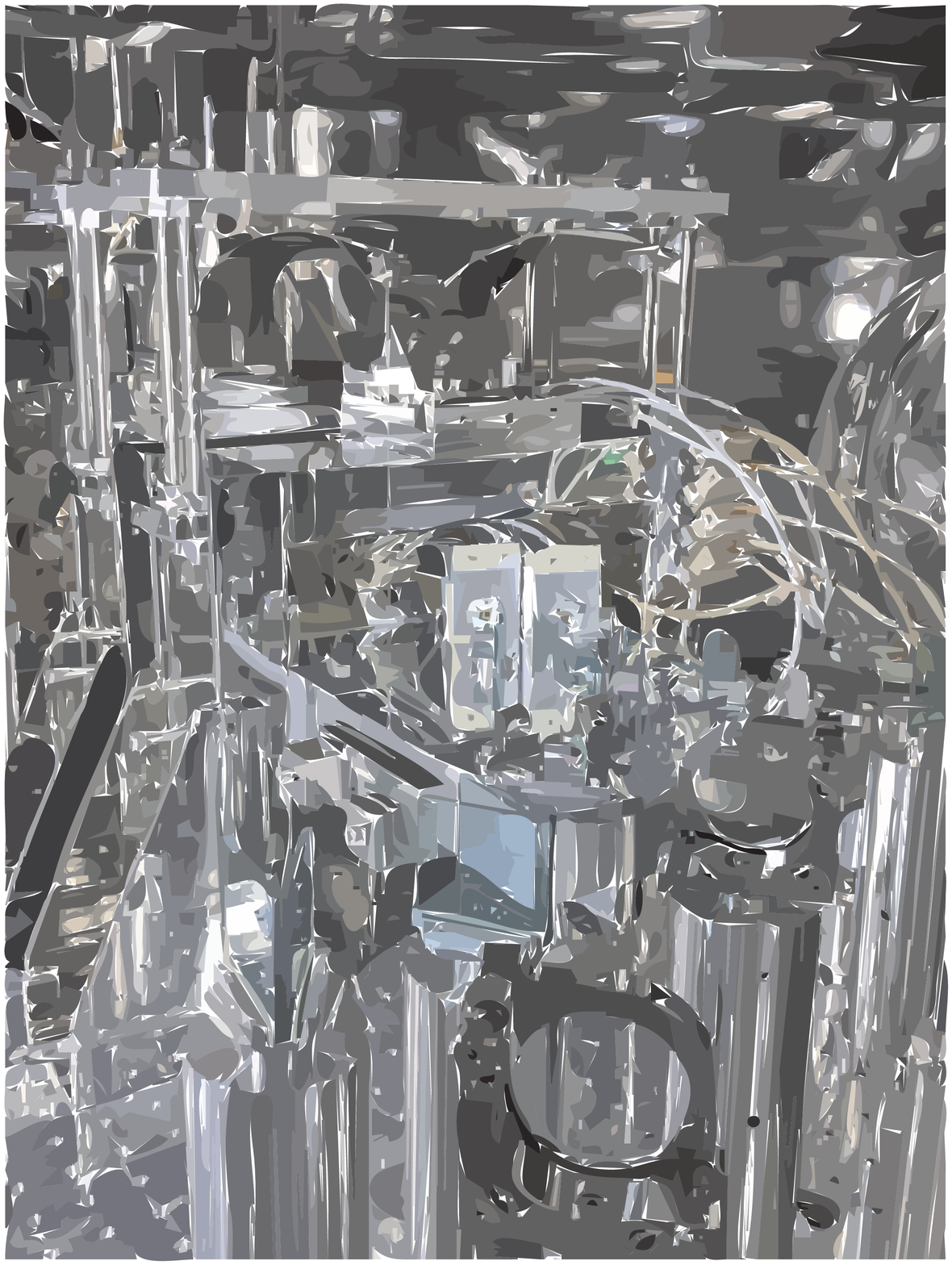}
	\caption{{\it Left}: Installation of the Type-A suspension system from the second floor of the Y-end station. {\it Center}: Assembly of the Type-Bp system outside the chamber. {\it Right}: Output mode cleaner installed in the vacuum chamber.}
	\label{fig:photos}
	\end{center}
\end{figure}

The 4-stage cryogenic payload is suspended by a 5-stage seismic isolation system, called Type-A (right panel of Fig.~\ref{fig:isolation}), at the room temperature in vacuum. The cryogenic payload and the Type-A isolation system is connected by a single maraging steel wire. The first stage isolation of the Type-A system consists of an inverted pendulum for horizontal isolation and a geometrical anti-spring (GAS) filter for vertical isolation. The inverted pendulum legs stand on the second floor of the central/corner station, which is approximately 15\,m above the ground floor of the underground facility. There are four more stages of GAS filters that undergo a bore hole between the second and ground floors. The Type-A system is 13.5\,m tall. The seismic attenuation rate was measured to be below $10^{-5}$ at 1.7\,Hz, which resulted in an attenuation rate of $10^{-21}$ at 10\,Hz for a carefully modeled suspension chain~\cite{OkutomiPhD}. The left panel of Fig.~\ref{fig:photos} illustrates the installation process of the Type-A suspension system from the second floor of the Y-end station.

The beam splitter and signal recycling mirrors are suspended by a 2-stage payload at the room temperature under a 3-stage seismic isolation system that includes inverted pendulums, called Type-B~\cite{TypeB}. The legs of the Type-B inverted pendulums stand on the frame built outside the vacuum chamber. The power recycling mirrors are suspended by a 3-stage isolation system without inverted pendulums, called Type-Bp~\cite{TypeBp}. The Type-Bp system is fixed on the frame built inside the vacuum chamber and is supported by a set of motorized linear stages to adjust the position and alignment of the system in the horizontal plane (see the center panel of Fig.~\ref{fig:photos}). The input mode-cleaner mirrors, the mode-matching telescope mirrors, and the output steering mirror are suspended by a double pendulum that was developed for test mass suspension in the TAMA300 telescope, called Type-C~\cite{TypeC}. The Type-C systems are fixed on three-stage vibration isolation stacks in the vacuum chambers.

Various types of actuators and local sensors are integrated in each suspension system for position and alignment control. In the GAS filter stages, linear variable differential transformers (LVDTs) are equipped to measure and control the vertical motion. In the top filter stages of the Type-A and Type-B suspension systems, a set of geophones and LVDTs for horizontal motion are also equipped. In the payload stages, coil-magnet actuators are attached to control the position and alignment of the mirror~\cite{Actuator}. The alignment of the suspended mirrors are monitored with optical levers and by optical shadow sensing of actuation magnets~\cite{OSEM}.

\subsection{Laser and input optics}
The laser source at the observing run in 2020 is a 400\,mW Nd:YAG NPRO laser amplified with a commercial fiber amplifier that yields an output power of 40\,W. 
In the near future, we will adopt a commercial solid-state amplifier 
to further increase the laser power. 
The laser light goes through a pre-mode cleaner to suppress the amplitude and phase noise in the radio-frequency band; subsequently, it goes through the modulation system to generate control sidebands at 16.875 and 45\,MHz~\cite{YamaKoh}. After exiting the laser room, the beam goes through the input mode cleaner with a round-trip length of 53.3\,m and finesse of 540.
The sideband frequencies are selected to be integer multiples of the free spectral range of the mode cleaner, namely 5.625\,MHz. Next the beam goes through the in-vacuum input Faraday isolator and the input mode-matching telescope; then, it enters the power recycling cavity. 
The input beam is s-polarized. The laser frequency is first stabilized for the input mode cleaner; it is then stabilized for the common mode motion of the 3\,km cavities. 
A rigid reference cavity is prepared in the laser room for pre-stabilizing the laser frequency, which is currently not used.
The laser intensity is stabilized using a transmitted beam of the first mode-matching telescope mirror (approximately 0.3\% transmission). The photo detector of the intensity stabilization system is currently located outside the vacuum and we plan to move it inside in the near future.

\subsection{Output optics}
The beam at the anti-symmetric port contains (i) gravitational wave signal sideband fields, (ii) 16.875\,MHz control sideband fields, (iii) higher order spatial modes of the carrier light, and (iv) a small fraction of carrier light that is to be used as a reference field for the detection of gravitational waves. Transmitting through the signal recycling mirror, the beam is sent to the output mode-matching telescope, which also helps in aligning the beam to the output mode cleaner with another steering mirror before the output mode cleaner. The beam then goes through the output Faraday isolator, which consists of a commercial Faraday crystal, a pair of Brewster angle polarizing beam splitters, and two quartz rotators. After exiting the output isolator, the beam is in the p-polarized state. The beam is reflected once by the suspended output steering mirror and then sent to the output mode cleaner, which is a 1.5\,m long, 4-mirror bow-tie cavity with a finesse of approximately 800. The cavity is mounted on an aluminum breadboard and the entire system is suspended by a single pendulum (see the right panel of Fig.~\ref{fig:photos})~\cite{KAGRAOMC3}. Unnecessary fields, namely (ii) and (iii) mentioned above, are filtered out by the output mode cleaner. The transmission of the fundamental mode carrier light measured outside the vacuum chamber before the installation is 85\%. The transmitted beam is split into two on-board photo-detectors ({\it Excelitas Technologies C30665GH}) and the sum of the signals is used as the gravitational wave channel.

\subsection{Digital control and calibration}
The control signals are obtained at numerous ports using photo-detectors and quadrant photo-detectors~\cite{AsoPRD}. The analog signals are converted to digital signals. The interferometer is operated using a digital real-time control system with high-speed front-end computers and a reflective memory system to clone data among distant computers. All computers are synchronized to the Global Positioning System (GPS) with negligible delay ($<1\,\mu\mathrm{s}$). 

It is essential to characterize the sensor and actuators to accurately estimate the amplitude and phase of the gravitational waves. We used two independent methods to improve the accuracy. In the first method, a photon calibrator was used to inject two 1047\,nm laser beams into different end test mass (ETM) locations and they were driven by the radiation pressure ~\cite{InoueGcal,PTEP03}. The ETM motion can be estimated from the power of the injected beams. The photon calibrator beam is injected from the vacuum chamber 36~m away from the ETM, because the view angle is narrow due to cryogenic duct shields.

The second method consisted of a so-called {\it free swinging Michelson method} in which the beam splitter was driven to observe several fringes. The fringe amplitude is calibrated to the displacement via the wavelength of the laser. Then, the actuator efficiency of the beam splitter was calibrated in terms of displacement per applied voltage as a function of frequency. Finally, using various interferometer configurations, the actuator efficiency of other optics were calibrated against that of the beam splitter by measuring transfer functions from these actuators to the interferometer output.
We used the photon calibrator as a primary tool and the free swinging Michelson method for cross-checking. 

The data were transferred from the KAGRA site to the data analysis building located nearby through a 7\,km optical fiber within 1 s, and then to the Institute for Cosmic Ray Research (ICRR), Kashiwa, in appoximately 3 to 10\,s. The gravitational-wave channel data were quickly transferred as low-latency data to Osaka City University in approximately 3\,s and LIGO/Virgo in approximately 9 to 15\,s. The bulk data were distributed to our Tier-1 sites at Academia Sinica (AS) in Taiwan and Korea Institute of Science and Technology Information (KISTI) in Korea in one day and then to Tier-2 sites in various places in a few days~\cite{KandaDMG,bKAGRAphase1}.

\section{Sensitivity}\label{ptep01_sec2}
In this section, we describe three fundamental noise sources that determine the target sensitivity of KAGRA, namely (i) quantum noise, which stems from the quantum fluctuation properties of light (photons); (ii) thermal noise, which is caused by the random actions of the heat bath; and (iii) seismic noise and seismic Newtonian noise, which originates from small ground vibrations that shake the mirrors via the vibration isolation system and Newtonian gravitation. We also explain how the current interferometer design was selected.

\subsection{Quantum noise}

The quantum noise of the gravitational-wave telescope originates from the zero-point fluctuations of the laser field. The Michelson interferometer is operated at the dark fringe where almost all of the incident laser field is reflected back toward the light source.
Therefore, the classical and quantum fluctuations accompanying the laser light do not leak to the anti-symmetric port. 
The vacuum fluctuations entering from the anti-symmetric port exclusively contribute to the sensitivity; the mean amplitude is zero but the mean square fluctuation is non-zero.

The quantum noise level can be derived by calculating the field propagation of the vacuum fluctuations~\cite{BC2001}. Correspondingly, the noise spectrum density is given by
\begin{eqnarray}
\sqrt{S_x(\Omega)}&=&\frac{h_\mathrm{SQL}}{\sqrt{2\cal K}}\frac{\sqrt{(A_{11}\cos{\zeta}+A_{21}\sin{\zeta})^2+(A_{12}\cos{\zeta}+A_{22}\sin{\zeta})^2}}{D_1\cos{\zeta}+D_2\sin{\zeta}}
\end{eqnarray}
where $\zeta$ denotes the readout phase (see Section~\ref{sec:sensitivity}),
\begin{eqnarray}
A_{11}&=&A_{22}\nonumber\\
&=&(1+r_s^2)\left(\cos{2\phi_\mathrm{SR}}+\frac{\cal K}{2}\sin{2\phi_\mathrm{SR}}\right)-2r_s\cos{[2(\beta+\Phi_\mathrm{SR})]},\nonumber\\
A_{12}&=&-t_s^2(\sin{2\phi_\mathrm{SR}}+{\cal K}\sin^2{\phi_\mathrm{SR}}),\nonumber\\
A_{21}&=&t_s^2(\sin{2\phi_\mathrm{SR}}-{\cal K}\cos^2{\phi_\mathrm{SR}}),\label{eq:inputoutput}\\
D_1&=&t_s(1+r_se^{2i(\beta+\Phi_\mathrm{SR})})\sin{\phi_\mathrm{SR}},\,\mathrm{and}\nonumber\\
D_2&=&t_s(1-r_se^{2i(\beta+\Phi_\mathrm{SR})})\cos{\phi_\mathrm{SR}}.\nonumber
\end{eqnarray}
Here, $r_s$ and $t_s$ denote the amplitude reflectivity and transmittance of the signal recycling mirror, respectively, $\phi_\mathrm{SR}$ is the phase rotation gained by the carrier in the signal recycling cavity ($\phi_\mathrm{SR}=[\omega_0\ell/c]_\mathrm{mod 2\pi}$), and $\Phi_\mathrm{SR}$ is the phase rotation gained by a signal sideband in the signal recycling cavity ($\Phi_\mathrm{SR}=[\Omega\ell/c]_\mathrm{mod 2\pi}$), wherein $\ell$ is the signal recycling cavity length and $\omega_0$ and $\Omega$ denote the angular frequency of the laser light (carrier) and gravitational wave (sideband), respectively. In addition, ${\cal K}$ is the optomechanical coupling coefficient of the Fabry-Perot Michelson interferometer, $\beta$ is the phase rotation gained by the signal sideband in the arm cavity, and $h_\mathrm{SQL}$ is the standard quantum limit (SQL) of the strain measurement of the interferometer.
\begin{eqnarray}
{\cal K}=\frac{8\omega_0I_0}{mL^2\Omega^2(\gamma^2+\Omega^2)},\,\ \tan{\beta}=\frac{\Omega}{\gamma},\,\ h_\mathrm{SQL}=\sqrt{\frac{8\hbar}{m\Omega^2L^2}}
\end{eqnarray}
where $I_0$ is the laser power at the beam splitter, $L$ is the arm cavity length, $m$ is the mass of the mirror, $\hbar$ is the reduced Planck constant, and $\gamma=\tau^2c/4L$ is the cavity pole angular frequency with $\tau$ being the amplitude transmittance of the input mirror.

\subsection{Thermal noise}
Thermal noise originates from statistical mechanics. The fluctuation-dissipation theorem predicts that the amplitude of the thermal fluctuation force is proportional to the product of temperature and dissipation~\cite{Levin}:
\begin{eqnarray}
P(\Omega) \propto T \phi,
\end{eqnarray}
where $P(\Omega)$, $T$, and $\phi$ denote the power spectrum density of the fluctuation, temperature, and loss angle, which represents the interaction between the system and heat bath, respectively.

LIGO and Virgo use fused silica as the material for the mirrors and suspension fibers.
A fused silica mirror ensures exquisite optical properties for well-developed $1\,\mu$m lasers and an extremely low mechanical loss, $\phi$, at the room temperature. 
Thermal noise can also be reduced by decreasing the operating temperature, $T$; this strategy was adopted in KAGRA. Fused silica, however, cannot be used as the material because its mechanical dissipation increases at the cryogenic temperature; therefore, sapphire crystals were used as the material for mirrors and suspension fibers in KAGRA.

\subsubsection{Mirror}

Dissipation sources that contribute to the mirror thermal noise are located in the substrate and coatings. Both exhibits two types of dissipation, i.e., structure damping and thermoelastic damping. Structure damping is empirically known to be almost independent of the frequency. Thermoelastic damping is caused by heat relaxation in the inhomogeneous strain.
Thermal noise caused by structure damping is called Brownian noise in this paper and that caused by thermoelastic damping is called thermoelastic noise. The power spectrum densities are given by~{\cite{Levin}\cite{Cerdonio}}
\begin{eqnarray}
P_{\rm Mirror sub Brown}(\Omega) &=& \left(\frac{2}{L}\right)^2 \frac{4 k_{\rm B}T (1-\sigma_{\rm sub}^2)\phi_{\rm sub}}{\sqrt{\pi}E_{\rm sub} w_0 \omega} \\
P_{\rm Mirror sub thermo}(\Omega) &=& \left(\frac{2}{L}\right)^2 \frac{16 k_{\rm B}T^2 \alpha_{\rm sub}^2 (1+\sigma_{\rm sub})^2w_0J_{\rm sub}(\Omega_c)}{\sqrt{\pi}\kappa_{\rm sub}}\\
J_{\rm sub}(\Omega_c) &=& \frac{\sqrt{2}}{\pi^{3/2}}\int_{0}^{\infty} du \int_{-\infty}^{\infty} dv \frac{u^3 {\rm e}^{-u^2/2}}{(u^2+v^2)[(u^2+v^2)^2+\Omega_c^2]}\\
\Omega_c &=& \frac{\Omega \rho_{\rm sub} C_{\rm sub} {w_0}^2}{2 \kappa_{\rm sub}}
\end{eqnarray}
The parameters $\sigma_{\rm sub}$, $E_{\rm sub}$, $\phi_{\rm sub}$, $\alpha_{\rm sub}$, $\kappa_{\rm sub}$, $\rho_{\rm sub}$, and $C_{\rm sub}$ represent the Poisson ratio, Young's modulus, loss angle of mirror substrate, thermal expansion coefficient, thermal conductivity, density, and specific heat per unit mass of the mirror substrate, respectively. The parameters $w_0$, $L$, and $k_{\rm B}$ denote the beam radius at mirror, arm length, and Boltzmann constant, respectively.

Although the reflective coating is four orders of magnitude thinner than the mirror substrate, its Brownian noise overcomes that of the substrate because the structure damping of the coating is four orders of magnitude larger than that of the substrate and the dissipation source is very close to the location of the probe~\cite{Levin}. 
The thermo-optic noise, which is a type of thermoelastic noise, is caused by the heat relaxation between the substrate and coating. The formulae of these noise sources are as follows~{\cite{Coating}\cite{CoatingTO}}:
\begin{eqnarray}
G_{\rm Mirror coating Brown} &=& \left(\frac{2}{L}\right)^2 \frac{8 k_{\rm B}T (1+\sigma_{\rm sub})(1-2\sigma_{\rm sub})d_{\rm coating}\phi_{\rm coating}}{\sqrt{\pi}E_{\rm sub} {w_0}^2 \omega} \\
G_{\rm mirror coating thermo} &=& \left(\frac{2}{L}\right)^2 \frac{2 k_{\rm B}T^2[2 \alpha_{\rm eff}d_{\rm coating}(1+\sigma_{\rm sub})-\beta_{\rm eff}\lambda]^2}{\pi w_0 \kappa_{\rm sub}}J_{\rm coating}(\Omega_c)\\
J_{\rm coating}(\Omega_c) &=& \frac{2\sqrt{2}}{\pi} \int_{0}^{\infty} du \int_{-\infty}^{\infty} dv \frac{u(u^2+v^2) {\rm e}^{-u^2/2}}{(u^2+v^2)^2+\Omega_c^2} 
\end{eqnarray}
where $d_{\rm coating}$ and $\phi_{\rm coating}$ denote the thickness and loss angle of the coatings, $\alpha_{\rm eff}$ and $\beta_{\rm eff}$ are the effective thermal expansion and effective temperature dependence of the refractive index of the coating, respectively, and $\lambda$ is the wavelength of laser. The structure damping of the coating depends on the temperature; it exhibits a gentle peak at approximately $20\sim30$\,K after annealing. As reported in a study conducted in Glasgow, the peak is sharp if the annealing temperature is high~\cite{Martin2008}. Such a sharp peak was not observed in our measurements in 2006~\cite{KYamamoto2006} and 2014~\cite{Hirosecoating}. 

\begin{figure}[htbp]
	\begin{center}
		\includegraphics[width=5cm]{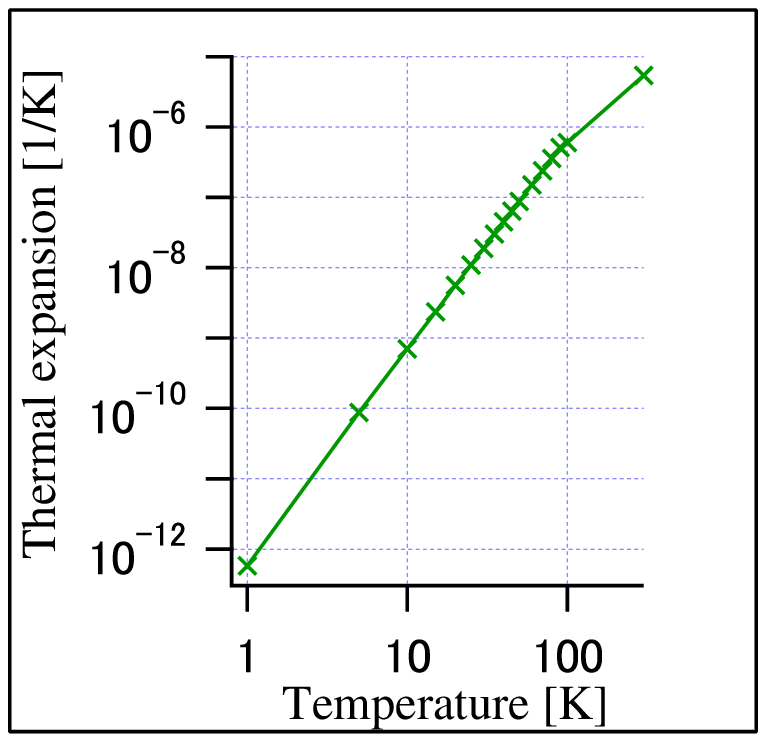}
		\includegraphics[width=5cm]{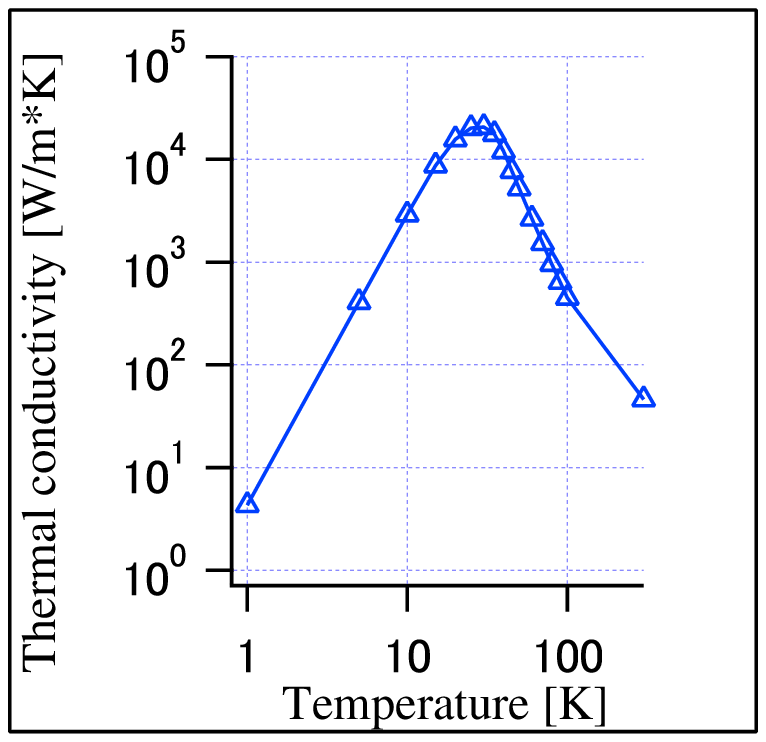}
		\includegraphics[width=5cm]{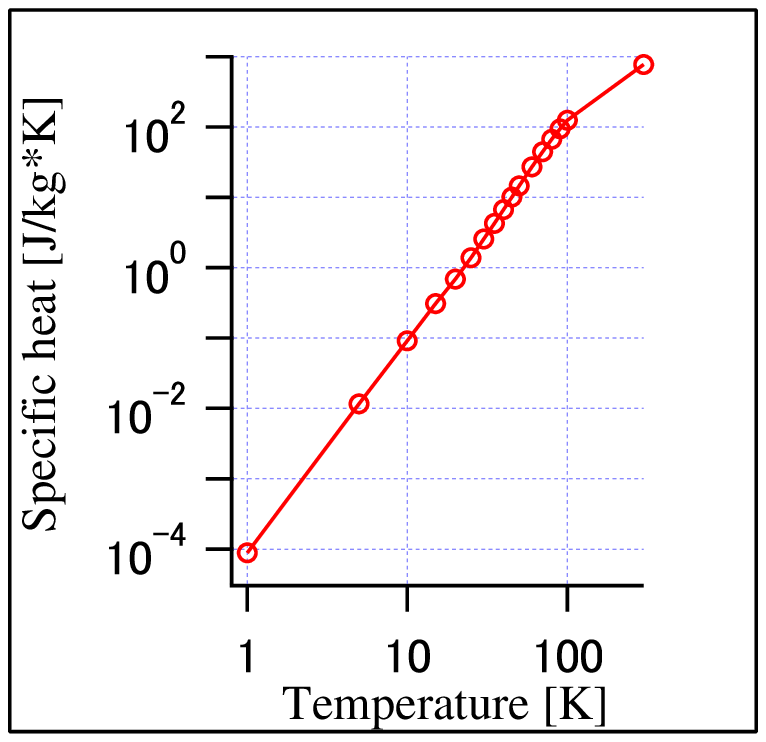}
	\caption{Thermal expansion ({\it left}), thermal conductivity ({\it center}), and specific heat ({\it right}) of sapphire as a function of temperature.}
	\label{fig:sapphireproperties}
	\end{center}
\end{figure}

The thermoelastic noise and thermo-optic noise depend on the thermal properties of sapphire. These properties, namely thermal expansion, thermal conductivity, and specific heat, vary significantly with temperature (see Fig.~\ref{fig:sapphireproperties}). Consequently, the thermal noise level of KAGRA varies significantly with its operating temperature at $T\gtrsim30$\,K, where the thermoelastic noise of the substrate exceeds the coating Brownian noise.

\subsubsection{Suspension}

We can focus on the suspension thermal noise of the bottom stage because the mechanical dissipation of the upper stages does not make significant contributions to the mirror motion~\cite{Saulson}.
A simplified expression of the power spectrum density of the thermal noise in its pendulum mode is given by
\begin{eqnarray}
P_{\rm thermal pendulum}(\Omega) = \left(\frac{2}{L}\right)^2 \frac{4k_{\rm B}Tg}{m \Omega^5\ell_{\rm fiber}}(\phi_{\rm fiber}+\phi_{\rm fiberthermo})\times\sqrt{\frac{4\pi E_{\rm fiber}}{mg\ell_{\rm fiber}^2}}\left(\frac{d_{\rm fiber}}{4}\right)^2,\label{eq:suspTN}
\end{eqnarray}
where $g$ is the gravitational acceleration, $\ell_{\rm fiber}$, $E_{\rm fiber}$, and $d_{\rm fiber}$ are the length, Young's modulus, and diameter, respectively, and $\phi_{\rm fiber}$ and $\phi_{\rm fiberthermo}$ denote the loss angles by structure damping and thermoelastic damping of the fiber, respectively.
To calculate the sensitivity curve, a more complicated equation based on the three-mass thermal noise model~\cite{PPP} is used and the temperature gradient in the sapphire fibers to extract heat is taken into account~\cite{KomoriSuspension}.

The last term in the right-hand-side of Eq.~(\ref{eq:suspTN}) depicts the ratio from its elastic restoring force to the gravitational force, which represents the dilution of the dissipation inside the suspension fiber~\cite{Saulson}. The suspension thermal noise can be reduced by using thinner fibers. The diameter of a fused silica fiber in a telescope at the room temperature is determined by the yield strength to be as thin as 0.5\,mm. The diameter of the sapphire fiber in KAGRA, however, is determined by the heat conductance to cool the mirror; it is as thick as 1.6\,mm. Thus, the dilution factor of the suspension KAGRA is almost unity. 

The thermoelastic damping in the fiber is given by~\cite{Saulson}
\begin{eqnarray}
\phi_{\rm fiber thermo} &=& \frac{\alpha_{\rm fiber}^2 E_{\rm fiber} T}{\rho_{\rm fiber} C_{\rm fiber}}\frac{f/f_0}{1+(f/f_0)^2}\,,\,\,\,\,
f_0 = 2.16 \frac{\kappa_{\rm fiber}}{\rho_{\rm fiber} C_{\rm fiber} d_{\rm fiber}^2} 
\end{eqnarray}
where $\alpha_{\rm fiber}$, $\rho_{\rm fiber}$, $C_{\rm fiber}$, and $\kappa_{\rm fiber}$ are the thermal expansion coefficient, density, specific heat per unit mass, and thermal conductivity of fibers, respectively. 
{\color{black}The thermoelastic damping of a sapphire fiber is several orders of magnitude smaller than the structure damping in the observation band at the cryogenic temperature. It is worth noting that} the temperature dependence of the Young's modulus of the fiber, which is used to cancel thermoelastic damping in LIGO or Virgo~\cite{Cagnoli}, is negligible for the sapphire fiber at the cryogenic temperature.
\begin{figure}[htbp]
	\begin{center}
		\includegraphics[width=15cm]{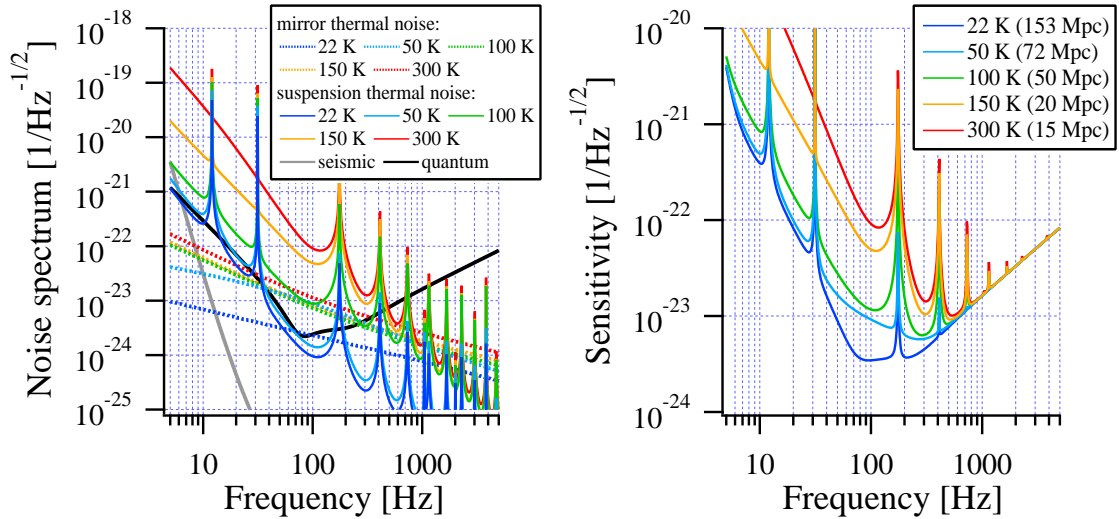}
\caption{{\it Left}: Mirror thermal noise and suspension thermal noise at 22, 50, 100, 150, and 300\,K. Seismic noise and quantum noise are also depicted. {\it Right}: Sensitivity curve at 22, 50, 100, 150, and 300\,K. The observation range of neutron star binaries at each temperature is shown in the legend.}
	\label{fig:mirrorTN}
	\end{center}
\end{figure}

The left panel of Fig.~\ref{fig:mirrorTN} shows the power spectra of the mirror thermal noise and suspension thermal noise at 22, 50, 100, 150, and 300\,K. It is evident that the mirror thermal noise increases rapidly between 22 and 50\,K, while the suspension thermal noise increases rapidly between 100 and 150\,K. This is because the thermoelastic noise increases above a certain temperature, which is given by the relaxation time of the temperature gradient in the material with the characteristic length. The characteristic length of the mirror thermal noise is determined by the beam radius and that of the suspension thermal noise is determined by the thickness of the suspension fiber; therefore, they increase rapidly at different temperatures. The right panel of Fig.~\ref{fig:mirrorTN} shows the sensitivity of the interferometer and the observation range of the neutron star binaries at each temperature.

\subsection{Seismic and seismic Newtonian noise}

\begin{figure}[t]
  \begin{center}
   \includegraphics[width=71mm]{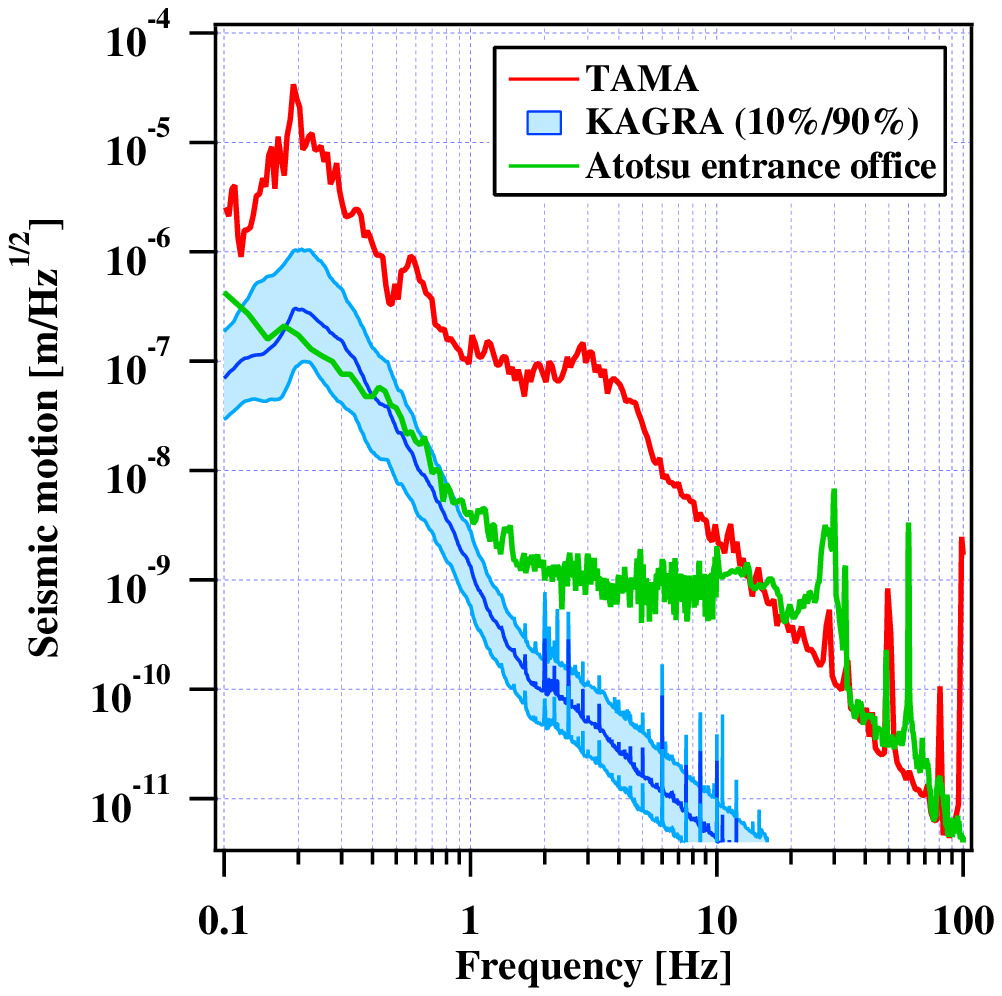}
   \includegraphics[width=71mm]{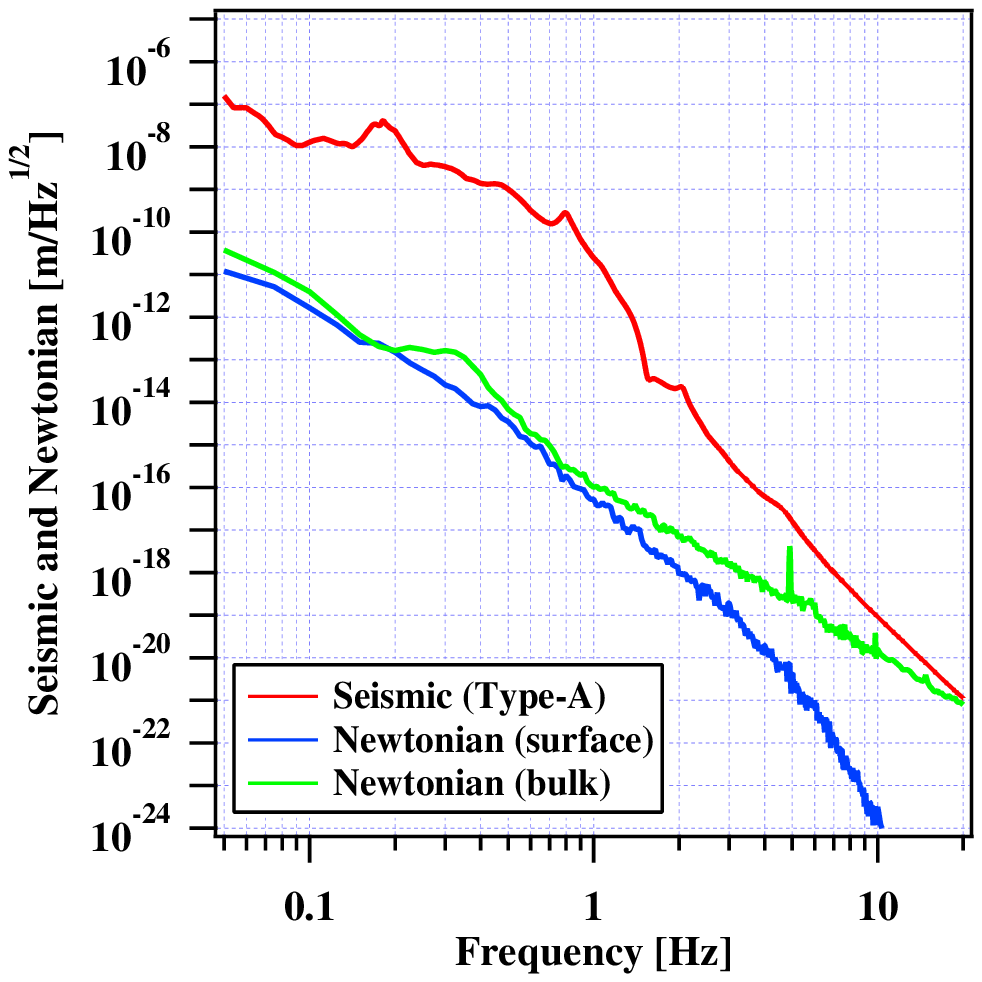}
  \end{center}
\caption{\label{fig:outside}{\it Left}:
Seismic motion spectra measured at the TAMA300 site in Tokyo, at a tunnel entrance near KAGRA (outside the mine), and at the KAGRA site (inside the mine). The 10\,\% and 90\,\% percentiles are also shown for the seismic motion at the KAGRA site. Below a few Hz, the seismic noise level is low in Kamioka even if it is outside the mine. Above 10\,Hz, however, the seismic noise level outside the mine is not significantly different from the one in Tokyo. The result indicates the importance of building a telescope inside the mine. {\it Right}: Seismic noise of the test mass suspended from the Type-A system, calculated using a suspension model and the measured seismic motion, along with the seismic Newtonian noise, which is calculated with the measured seismic motion. The Newtonian noise from the ground surface is suppressed under the ground.}
\end{figure}

Because KAGRA is located deep under the ground, the level of seismic noise observed at its site is significantly lower than that of other second generation gravitational-wave telescopes. A preliminary survey proved that the seismic noise level is low (i) below 1 Hz for being well isolated from human activities and (ii) above 10 Hz for being built under the ground, as shown in the left panel of Fig.~\ref{fig:outside}~\cite{iKAGRA:PTEP}. The difference between the levels of seismic noise inside and outside the mine is approximately a factor of 100 or more above 10\,Hz.

Another advantage of KAGRA is its low seismic Newtonian noise. There are two types of seismic Newtonian noise; one originates owing to the vibrations in the ground surface mainly due to human activities and the other originates owing to the bulk motion of the Earth. In telescopes that are built on the ground or near the surface, the former type of Newtonian noise is more prominent. In KAGRA, the bluk noise effect is the same as that of other telescopes but the surface noise effect is suppressed (see the right panel of Fig.~\ref{fig:outside}). In comparison to LIGO, for example, the total seismic Newtonian noise level is approximately one order of magnitude smaller above 10\,Hz.

\subsection{Design sensitivity}\label{sec:sensitivity}

\begin{figure}[t]
  \begin{center}
   \includegraphics[width=80mm]{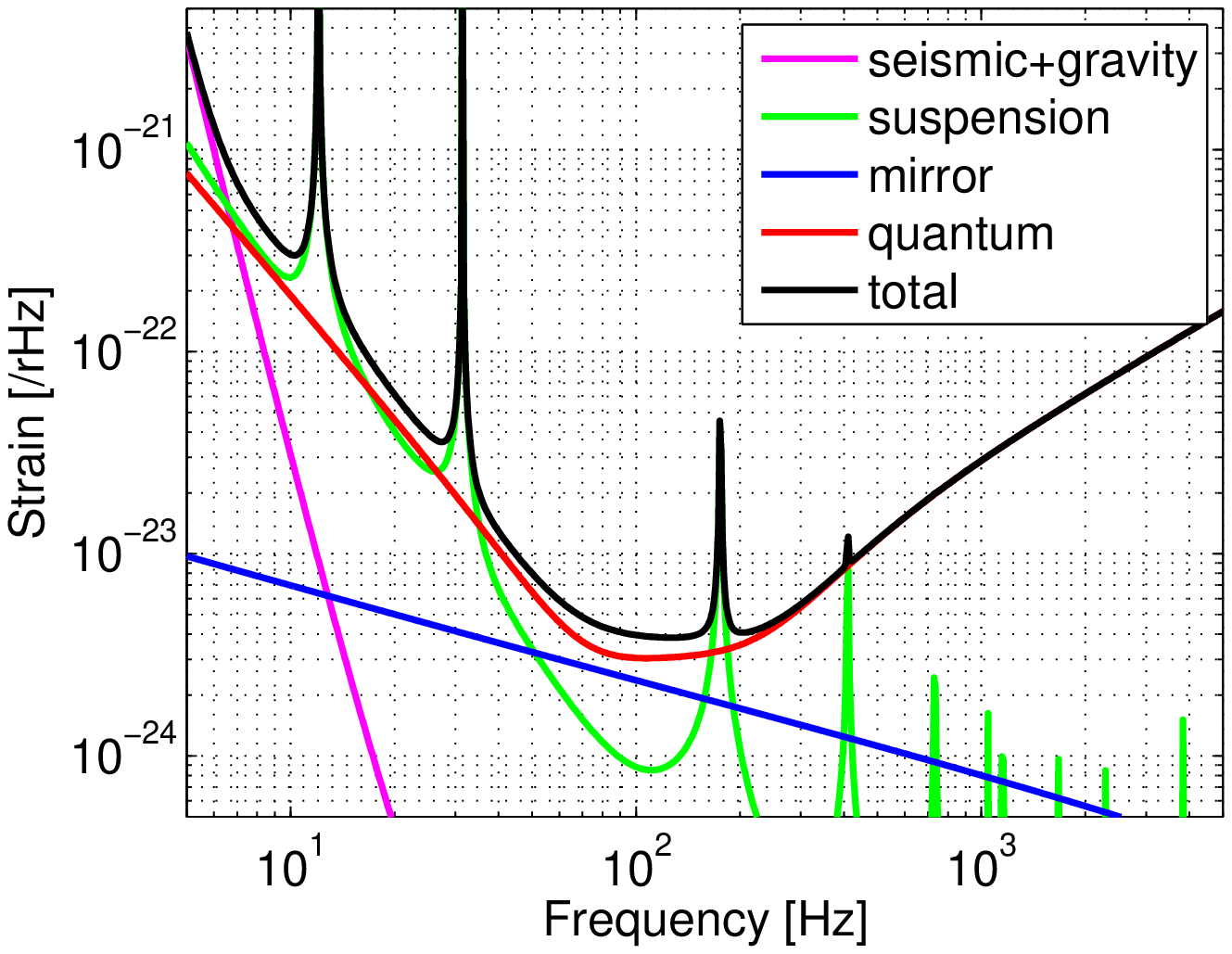}
   \raisebox{7mm}{\includegraphics[width=70mm]{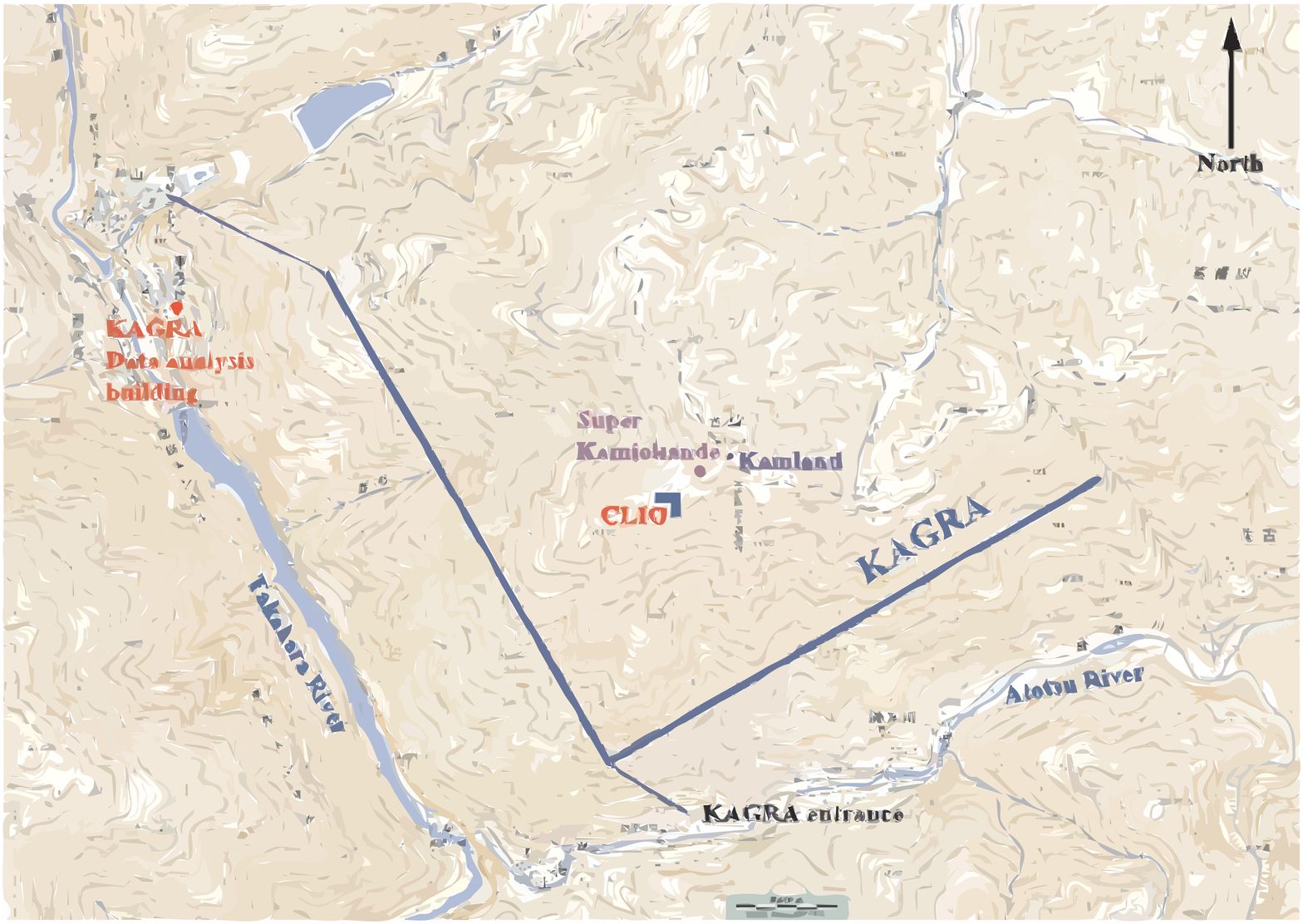}}

  \end{center}
\caption{\label{fig:kando}{\it Left}: Design sensitivity of KAGRA. {\it Right}: A map of the KAGRA site~\cite{iKAGRA:PTEP}.}
\end{figure}

As shown in the left panel of Fig.~\ref{fig:kando}, the design sensitivity of KAGRA is determined by the squared sum of the fundamental noises described in the previous sections. Some parameters were slightly modified during the installation phase but the design sensitivity is almost the same as the one introduced in Ref.~\cite{SomiyaCQG}. The parameters for the design sensitivity are summarized in Refs.~\cite{officialsensitivity,MichimuraPSO} and are listed in Table~\ref{table:parameters}.

\begin{table}[h]
\begin{center}
\begin{tabular}{lc}
\hline
Item & Parameters \\ \hline
Laser power at BS $I_0$& 674 W \\
Power transmittance of ITM $\tau^2$& 0.004\\
Amplitude reflectivity of SRM $r_s$& 0.92\\
Round trip loss of arm & 100 ppm\\
Power loss at SRM & 0.002\\
Power loss at PD & 0.1\\
Detune phase $\pi/2-\phi$ in degrees & 3.5$^{\circ}$\\ 
Readout phase $\zeta$ in degrees & 135.1$^{\circ}$\\ \hline
Test mass radius & 11 cm \\
Test mass thickness & 15 cm \\
Mass $m$& 22.8 kg\\
Temperature $T$& 22 K\\
Loss angle of substrate & 1.0 $\times 10^{-8}$\\
Optical absorption of substrate & 50 ppm/cm\\ \hline
Number of coating layers & 22/40 (ITM/ETM)\\
Loss angle of silica coatings & 3.0 $\times 10^{-4}$\\
Loss angle of tantala coatings & 5.0 $\times 10^{-4}$\\
Beam radius (ITM/ETM) & 3.5 cm/3.5 cm\\
Optical absorption of coatings & 0.5 ppm\\ \hline
Number of suspension fibers & 4 \\
Suspension fiber length & 35 cm\\
Suspension fiber diameter & 1.6 mm\\
Loss angle of suspension fibers & 2 $\times 10^{-7}$\\
Average temperature of suspension fibers& 19 K (top: 16 K and bottom: 22 K) \\ 
Vertical-Horizontal Coupling & 1/200 \\ \hline
\end{tabular}
\caption{\label{table:parameters}List of important parameters of the KAGRA telescope used for calculating the design sensitivity. Here, ITM and ETM stand for input test mass and end test mass, respectively.}
\end{center}
\end{table}

The primary interferometer configuration comprises a detuned RSE. The arm cavity finesse, signal recycling gain, detune phase of the signal recycling cavity, and readout phase were selected to improve the quantum noise level at around 100\,Hz, where KAGRA has an advantage in comparison to other telescopes, without significantly sacrificing the sensitivity at other frequencies. Reducing the quantum noise results in the improvement of the observational range of the compact binaries by approximately 20\% in comparison to the non-detuned configuration. It should be noted that some technological challenges were considered in the selection process, i.e., the upper limit of the arm cavity finesse was determined to be 1550 according to a concern about the increasing loss in the signal recycling cavity, and the upper limit of the detune phase was determined to be $3.5^\circ$ owing to the control scheme of the signal recycling cavity~\cite{AsoLSC}.

The readout phase $\zeta$ was optimized to increase the observation range of the compact binaries. The current readout scheme uses a fraction of the carrier light leaking from the arm cavities for the imbalance in the mirror reflectivity in the two arms and for an offset added to the differential control of the arm cavities. The readout phase can be tuned by balancing the two components~\cite{SomiyaLaserNoise}. An alternative method has been proposed for other telescopes to realize the optimization in more flexible manner using a so-called balanced homodyne detector~\cite{BHD}.

\section{Construction history}\label{ptep01_sec4}

\subsection{Site search and tunnel excavation}

In the late 1990s, we consulted a company to survey several candidate sites in Japan, including Mt. Tsukuba (Ibaraki Pref.), Kamaishi mine (Iwate Pref.), and the Kamioka mine (Gifu Pref.). The Kamioka site was selected owing to its low estimated cost of excavation and some experiences of measurements near the neutrino detector Super-Kamiokande. The site is surrounded by the Hida Gneiss (tough, hard, coarse-grained metamorphic rock), the oldest rock in Japan. 

\begin{figure}[htbp]
	\begin{center}
		\includegraphics[height=3.8cm]{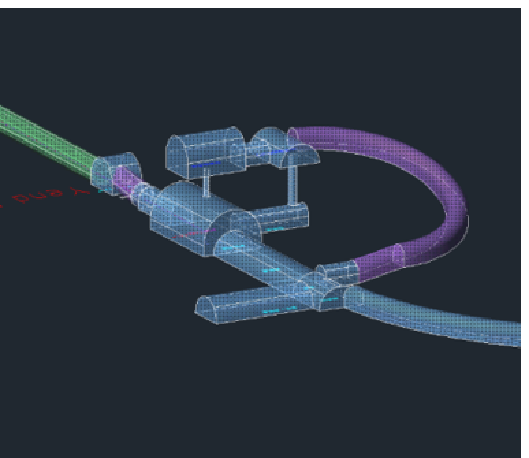}
		\includegraphics[height=3.8cm]{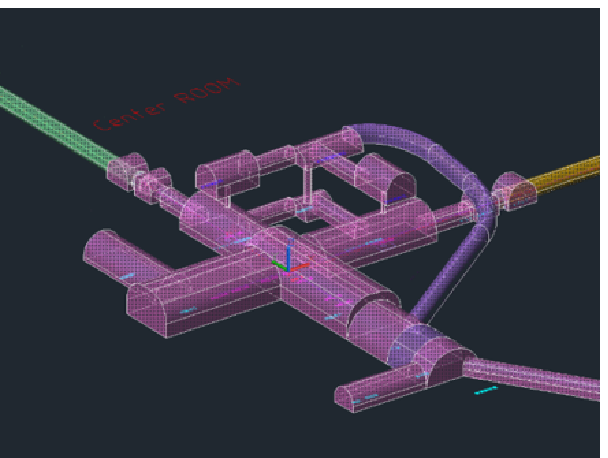}
		\includegraphics[height=3.8cm]{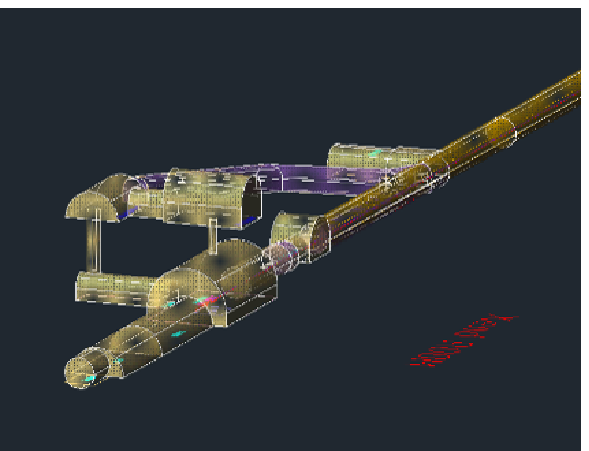}
	\caption{Three-dimensional CAD of the Y-end station ({\it left}), central station ({\it center}), and X-end station ({\it right}).}
	\label{fig:tunnel}
	\end{center}
\end{figure}

The KAGRA project was approved in 2010 with the starting budget 14 billion JPY. The tunnel excavation started in May 2012 and finished in March 2014
using the new Australian tunneling method~\cite{UchiyamaCQG}. The geographical coordinates of the beam splitter are 36.41$^{\circ}$ N and 137.31$^{\circ}$ E. The Y-arm is in the direction 28.31$^{\circ}$ NW. The distance from the ground surface to the underground KAGRA facility is 200\,m or more. There is an access tunnel to the central station and another access tunnel to the Y-end station (see the map in the right panel of Fig.~\ref{fig:kando}). Both arms are tilted by $1/300$ to drain the spring water. The X- and Y-end represent the highest and lowest points, respectively; approximately 1,200 tons of water at most per year passes through the drainage pipes near the Y-end station. 
The central station and the two end stations have the second floor to accommodate the Type-A suspension systems. Figure~\ref{fig:tunnel} shows three-dimensional CAD pictures of each station.

The installation of the clean rooms and vacuum chambers started in March 2014, parallel to the last fraction of the excavation work. In the beginning of the installation work, we found spring water on the floor and walls; therefore, we covered the walls with vinyl sheets and plastic paint and dug holes to release the water under the floor. After the treatments, the water issue was solved~\cite{Miyoki2019}. The amount of water was the highest in April 2015 (approximately $1250\,\mathrm{t/h}$ was drained at the end of the Y-arm tunnels) and it decreased in subsequent years. 

\subsection{Test operations}

\begin{figure}[t]
  \begin{center}
   \includegraphics[width=140mm]{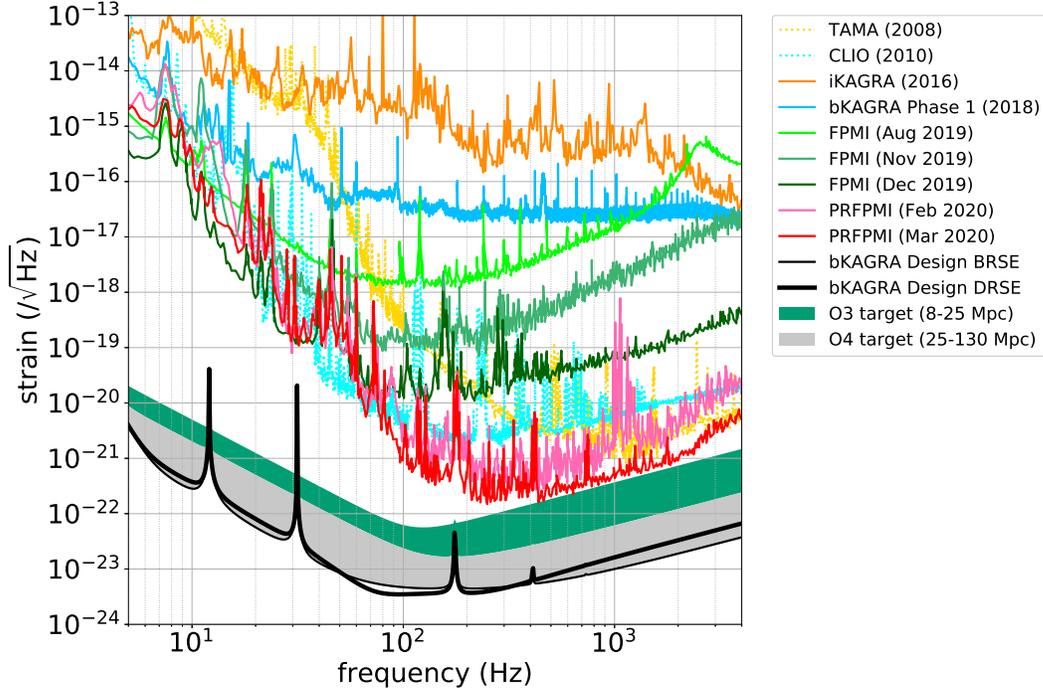}
  \end{center}
\caption{\label{fig:SensitivityHistory}
Sensitivity plots in the test operations and during the commissionings toward the observing run. Typical sensitivity curves during the iKAGRA and bKAGRA Phase 1 test operations are shown in orange and light blue, respectively. The first sensitivity curve with the Fabry-Perot Michelson interferometer (FPMI) configuration obtained in August 2019 is shown in light green. The best sensitivity with FPMI configuration obtained in December 2019 is shown in green. The sensitivity curve with the power-recycled Fabry-Perot Michelson interferometer (PRFPMI) configuration at the beginning of the observing run in February 2020 is shown in pink. The sensitivity obtained after the one month commissioning break is shown in red.  These sensitivity curves are obtained from the online calibration pipeline~\cite{PTEP03}. The green and grey shaded regions show the target sensitivities for O3 and O4, respectively~\cite{ObservingScenario}. Typical sensitivity curves of TAMA300~\cite{TAMA} and CLIO~\cite{CLIO} are also shown for comparison.
}
\end{figure}

In March 2016, after installing the most basic set of hardware, we performed a test operation on KAGRA as the first trial of controlling the 3 km interferometer in Japan~\cite{iKAGRA:PTEP, iKAGRAICA}. 
The interferometer configuration was selected to be a simple Michelson interferometer with silica test masses suspended by small suspensions similar to the Type-C suspension.
This configuration was called {\it iKAGRA}. 
The main goal of this operation was to ensure the overall performance of our first km-scale telescope and test the digital control system. The test operation was successful.
The noise level of the gravitational-wave channel is shown in the orange plot in Fig.~\ref{fig:SensitivityHistory}.

In April 2018, when the sapphire mirrors were almost ready, we performed a cryogenic test operation on the 3 km Michelson interferometer~\cite{bKAGRAphase1}. Owing to a delay in the installation process, only the X-end mirror (ETMX), kept at the room temperature, was used as the actual high-quality sapphire mirror in the final configuration. The Y-end mirror (ETMY), cooled below 20\,K, was a prototype mirror for the test.
This configuration was called {\it bKAGRA phase-1}. The main goal of the second test operation was to examine the performance of the full Type-A seismic isolation system that suspended the ETMX and to examine the performance of the cryogenic system for the ETM; both were successful. In 30 days, the ETMY was cooled to 18\,K. We also examined the KAGRA data transfer system for the first time. The measured data transfer speed was $20\,\mathrm{MB/s}$, which satisfied the requirement. The noise level of the gravitational-wave channel is depicted in the light blue plot in Fig.~\ref{fig:SensitivityHistory}.

\subsection{Toward the first observation run}

After the test operation of bKAGRA phase-1, we installed the input mirrors (ITMX and ITMY) and replaced the prototype ETMY with a high-quality mirror. 
The installation of all test masses was finished by November 2018. We also installed the remaining main suspensions such as the signal recycling mirror (SRM) by December 2018. For the SRM, we installed a mirror with a power reflectivity of 70\% instead of original 85\%. This choice was made to account for the lower input power to the interferometer in the first observing run~\cite{O3Sensitivity}. The 70\% SRM was a compound optics with a 2-inch fused silica optic mounted in an aluminum ring with a diameter of 100 mm.

The X- and Y-arm cavities were successfully controlled at the operation point ({\it locked}) for the first time in January 2019 and July 2019, respectively, using the arm length stabilization system described in Ref.~\cite{EnomotoALS}, and the Fabry-Perot Michelson interferometer (FPMI) was locked in August 2019. 
The mirrors were once cooled to $\sim20$\,K in 2019, but we stopped the cryogenic cooling during interferometer commissioning mainly due to a frosting issue. The X-arm finesse ($1410\pm30$ at the room temperature) was observed to be 1430 at test mass temperatures 29\,K (input) and 38\,K (end); however, it decreased down to 990 when the test mass temperature was 23\,K. The finesse 
 was restored to $1410\pm30$
after the input test mass temperature increased to 55\,K (the end test mass temperature was 83\,K). Similarly, the Y-arm finesse was measured to be below 500 when both test masses were cooled below 30~K, but it recovered to its normal value of $1330\pm20$ after both test masses were warmed above 60~K.

At this time, the vacuum level was on the order of $10^{-6}$\,Pa. When the test mass temperature was increased to approximately 30-33\,K, an outgassing peak was observed. Therefore, we concluded that the lower finesse was caused by the frosting of the residual gas molecules. 
We decided not to cool the payload, and to cool only the cryogenic duct-shields to 80\,K because they act as cryo-pumps to reduce the vacuum pressure.
The radiation from the test mass to the cryogenic duct-shield decreased the test mass temperature to around 250\,K during the commissioning. See Section~\ref{sec:challenges} for more details.


During the commissioning, we also found inhomogeneous birefringence in the input test masses (ITMs). A fraction of the s-polarized light that transmits through the ITMs scatters to the p-polarized light. The amount of power loss to p-polarization on reflection from ITMX and ITMY was measured to be 6.1\% and 11\%, respectively, when the arm-cavities were kept off-resonance. These measured losses were consistent with the values estimated from the transmission map with different polarizations~\cite{JGW-G1910369,JGW-G1910388}. Because this birefringence was inhomogeneous, the amplitude and phase of p-polarization varied with the spot positions of ITMs; this prevented the power-recycled Michelson interferometer from performing stable operations.

When the arm cavity was locked, however, the power of p-polarization inside the power recycling cavity was reduced by a factor of three, presumably from the cancellation effect. Because the sign of the incident field promptly reflected from the ITMs and that of the intracavity field transmitted through the ITMs are opposite, the effect of transmission wave front error was canceled for the carrier field. This effect has been recognized in LIGO for thermal lensing~\cite{LawrenceEffect}, and we found that this effect is also applicable to birefringence. See Section~\ref{sec:challenges} for more details.

Figure~\ref{fig:SensitivityHistory} shows the sensitivity improvement during the commissioning work before the observing run. The dark green curve shows our best sensitivity with the FPMI configuration achieved in December 2019. The first lock of the power-recycled Fabry-Perot-Michelson interferometer (PRFPMI) was achieved in January 2020. The output mode cleaner~\cite{KAGRAOMC3}, installed in 2018 and refurbished in 2019, was ready by the time, and we successfully upgraded the signal readout scheme from the conventional RF readout to the DC readout~\cite{DCreadout} in February. These enabled us to improve the sensitivity by increasing the circulation power in the interferometer. The power recycling gain for the carrier field in the PRFPMI configuration was measured to be $11\sim12$. 
This value is consistent with the originally designed value, and roughly agrees with the value estimated from the reflectivity of the arm cavities for s-polarization, considering the cancellation effect described above for the birefringence.
The floor level of the sensitivity curve has been improved by 3-4 orders of magnitude in six months and the observation range for the neutron star binaries has been improved by three orders of magnitude. The current sensitivity is mostly limited by control noise of various degrees of freedom of the interferometer at low frequencies, and it is limited by shot noise, laser intensity noise, and laser frequency noise at high frequencies. We note here that the tilted SRM in the PRFPMI configuration introduces an optical loss of 70\%.

Although the sensitivity is not comparable to that of Advanced LIGO or Advanced Virgo, KAGRA officially started its first observing run on February 25 (2020) with a maximum binary neutron star range of approximately 600\,kpc. On March 10, we temporarily halted the observation to perform a series of adjustments in the interferometer alignment and further investigations on the laser intensity and frequency noise coupling. After a month-long commissioning break, KAGRA achieved the maximum binary neutron star range of approximately 1\,Mpc, and resumed the observing run on April 7. While the sensitivity could not reach the original O3 target of 8-25 Mpc, as shown in Fig.~\ref{fig:SensitivityHistory}, we kept the interferometer in the observation mode until April 21.

\subsection{Challenges toward future observations}\label{sec:challenges}

In this sub-section, we introduce the challenges toward future observations, focusing on those that are unique to the underground cryogenic telescope.
\begin{itemize}
    \item {Newtonian noise of the spring water: Over 1200 t/h of spring water can be drained through the water pipe near the Y-end. According to the calculations performed by Chen and Nishizawa, the Newtonian noise of the spring water can limit the sensitivity at low frequencies, depending on its average velocity~\cite{waterGGN}. This is being further analyzed via a fluid dynamics simulation.}
    
 \item{Ice layers on the mirror surface: 
    A significant number of water or other molecules can get attached to the cryogenic mirror if the vacuum level is not high enough during the cooling process. The frosting issue discussed in the previous section is expected to be solved by properly cooling the shields. 
    {\color{black}It is important for the temperature of the sapphire mirror to be higher than that of the inner radiation shield. We should first cool the cryogenic duct-shields enough to achieve a good vacuum and then start cooling the payload}. 
    However, even under a good vacuum, the adsorbed molecules form adlayers that grows with time. Such adlayer formation will result in variations in the reflectivity of the test masses and also increase of thermal noise, as stated in Refs.~\cite{Hasegawa2019,Steinlechner2019}. 
    {\color{black}A detailed analysis of a method to remove such adsorbed molecules through photostimulated desorption is being conducted}}
    \item{Birefringence of sapphire: Inhomogeneity observed in the sapphire crystals used for the input mirrors is a big challenge. To ensure that the wavefront of the transmitted beam from the input test mass is uniform, optical thickness for the s-polarized light needs to be independent of the position on the test mass. The transmission wavefront error is as large as a few tens of nanometers for the 15 cm-thick crystal, but this error can be reduced to a few nanometers with the ion-beam figuring (IBF) treatment. Since sapphire has birefringence, the optical thickness difference between s-polarization and p-polarization also has to be uniform. This can be realized by aligning the sapphire c-axis to the beam propagation axis and aligning the ordinary axis with the polarization plane.
    
    When we manufactured the input test masses, we used circular polarization for the wavefront measurements instead of s-polarization to correct the error with IBF treatment. This resulted in the error measured with s-polarization to be approximately 30\,nm for both input mirrors while the requirement was 6\,nm~\cite{HiroseMirror}. According to our simulations, this resulted in several non-trivial problems~\cite{SomiyaITM}. We also didn't took much attention on aligning the sapphire axes to the beam, which resulted in inhomogeneous birefringence, as discussed in the previous section.
    
    To ensure that the power lost owing to the reflection is less than 1\%, the optical path length difference between two linear polarizations must be within 5\,nm. This is equivalent to aligning the sapphire c-axis to the beam propagation axis within 0.13$^{\circ}$~\cite{JGW-G1910369}. If the axes are well aligned, a precise IBF treatment using s-polarization will introduce a negligible birefringence effect on the interferometer. Research and development of a homogeneous sapphire crystal in terms of birefringence are being conducted within the collaboration.}

    \item{Low lifetime of the cryo-cooler: The lifetime of the rotary valve unit used for the cryogenic duct-shields is as short as 3,000 h. There are a total of eight units; therefore, the mean time between failure (MTBF) is only 15 days. We have been maintaining the cryo-cooler every 15 days but this issue must be solved before making longer observations in the future. Additionally, 16 cryo-cooler units are used for the cryostats (with a longer lifetime); their MTBF is 52 days. The MTBF of the compressors is 104 days for the cryogenic duct-shield and 52 days for the cryostats. We monitor the temperature of each duct-shield and exchange the valve when the temperature starts to increase. In the near future, we should replace the cryo-coolers at least for the duct-shields.} 
\end{itemize}

\section{Outlook}\label{ptep01_sec5}
In this paper, we reported the design and construction processes of the gravitational-wave telescope KAGRA. KAGRA started its first observing run in February 2020 and it can observe gravitational waves from neutron star binaries at a distance of approximately 1\,Mpc from the Earth with its latest sensitivity. The current sensitivity of this telescope is not as good as that of Advanced LIGO or Advanced Virgo, but we expect to achieve a comparable sensitivity within the next few years.
While the sensitivity has been significantly improved in the past six months, we also noticed several non-trivial problems, unique to underground cryogenic interferometers, that could prevent us from reaching the target sensitivity of approximately 150\,Mpc. 
The Addition of KAGRA to the global network of advanced telescopes can enable the precise estimation of parameters and accelerate further astrophysical discoveries. With underground and cryogenic technologies, KAGRA presents a pioneering future technology to realize the next generation of telescopes.


\section*{Acknowledgment}
This work was supported by the MEXT, JSPS Leading-edge Research Infrastructure Program, JSPS Grant-in-Aid for Specially Promoted Research 26000005, JSPS Grant-in-Aid for Scientific Research on Innovative Areas 2905: JP17H06358, JP17H06361 and JP17H06364, JSPS Core-to-Core Program A. Advanced Research Networks, JSPS Grant-in-Aid for Scientific Research (S) 17H06133, the Mitsubishi Foundation, the joint research program of the ICRR (University of Tokyo), National Research Foundation and Computing Infrastructure Project of KISTI-GSDC in Korea, AS, AS Grid Center, and the Ministry of Science and Technology in Taiwan under grants including AS-CDA-105-M06, the LIGO project, and the Virgo project. We would like to thank Editage (www.editage.com) for English language editing.

\end{document}